\documentclass{article}

\usepackage{PRIMEarxiv}

\usepackage[utf8]{inputenc} 
\usepackage[T1]{fontenc}    
\usepackage{hyperref}       
\usepackage{xurl}           
\usepackage{booktabs}       
\usepackage{amsfonts}       
\usepackage{nicefrac}       
\usepackage{microtype}      
\usepackage{fancyhdr}       
\usepackage{graphicx}       
\usepackage{subcaption}     
\usepackage{placeins}       
\graphicspath{{media/}}     

\makeatletter
\AtBeginDocument{\let\oldsection\section
  \renewcommand\section{\@ifstar{\oldsection*}{\FloatBarrier\oldsection}}}
\makeatother

\title{Target Speaker Identification: A Low-Latency Streaming Pipeline
}

\author{
  Patrick S. Burke \\
  Children's National Hospital \\
  Washington, DC \\
  \texttt{Pburke2@cnmc.org} \\
  \And
  Satyam Raj \\
  Arizona State University \\
  Tempe, AZ \\
  \texttt{sraj17@asu.edu} \\
  \And
  Sean Kinahan \\
  Arizona State University \\
  Tempe, AZ \\
  \texttt{skinahan@asu.edu} \\
}

\begin{document}
\maketitle

\begin{abstract}
We present a real-time pipeline of open source, pretrained models for streaming identification of a target speaker, motivated by hearing-aid applications where latency as low as 10 ms can be perceptible. We formulate a two-step approach in which incoming audio is first segmented by speaker using low-latency streaming diarization, followed by speaker verification against a registered target speaker. To emulate conversational speech while minimizing overlap, we use the This American Life Podcast Transcripts dataset and select the host as a consistent target speaker. We benchmark offline diarization with Pyannote and LIUM using diarization error rate (DER) and select Pyannote based on baseline performance and compatibility with streaming. We then evaluate speaker verification using Pyannote and TitaNet-Large and generate ROC curves to select an operating region. We integrate Diart and tune clustering parameters to reduce DER while maintaining real-time operation. We pair Diart with Pyannote verification and evaluate system-level performance by converting predicted and ground-truth speech regions into 100 ms binary masks. Across 17 evaluation episodes, the system achieves greater than 0.90 median accuracy with high specificity (0.95--0.98) at cosine distance thresholds of 0.7--0.75, demonstrating a practical proof of concept for downstream low-latency selective amplification.
\end{abstract}


\section{Introduction}

Hearing loss is among the most common sensory impairments, and nearly 70\% of adults have some level of hearing loss by age 71 \cite{reed2023prevalence}. It has been implicated as a risk factor for Alzheimer's disease, impaired physical function, and all-cause mortality \cite{reed2023prevalence,choi2024association,schuster2021conventional}. Additionally, 2--3 per 1000 children are born with some level of congenital hearing loss, which has been associated with speech and language delays, impaired executive function, and hyperactivity disorders \cite{lieu2020hearing}. Hearing aids are medical devices that process and amplify an audio signal before presenting it to a user's ears \cite{petersen2022effects}. Although hearing aid usage is somewhat low, with only about 30\% of adults with hearing loss wearing them, studies have found that the introduction of hearing aids can mitigate health \cite{reed2023prevalence} and developmental \cite{lieu2020hearing} concerns.

Significant challenges remain even for users of hearing aids. Amplified speech quality is degraded in settings with multiple speakers or background noise, which may lead to increased communication effort and hearing fatigue \cite{petersen2022effects,alhanbali2017self}. In the long term, this contributes to decreased social participation, loneliness, and the risk for social isolation \cite{jansen2025wham}. Although hearing aids have built-in noise reduction, nearly half of users are unhappy with their hearing aids' performance in noisy environments \cite{chen2021effects}. One common solution is the use of a portable microphone, although this approach may be impractical in many circumstances. For example, a student with hearing loss may set up a microphone during class, which requires setup time at the start of each class and keeping track of an additional accessory beyond the student's hearing aids. Furthermore, cost has been cited as a major barrier in purchasing compatible microphones \cite{scarinci2022qualitative}, which increases the risk for inequity of access to high-quality hearing assistance.

In order to address these concerns, we developed a real-time pipeline of open source, pretrained models for speaker diarization and verification that can be used for streaming identification of a target speaker and used in downstream applications such as audio processing within hearing aids. Here, a two-step approach is described in which audio is segmented according to speaker using low-latency streaming diarization, followed by speaker verification against a registered target speaker using machine learning (ML) models.

Speaker diarization is the partitioning of a continuous audio stream into discrete segments based on the identity of the speaker. More recent end-to-end diarization systems can handle overlapping speakers, an unknown number of speakers, and allow for streaming (or ``online'') audio \cite{park2022review,xue2021online}. Speaker verification is next required to identify a target speaker by comparing an unknown audio segment with a known target speaker audio segment to determine if they match. Throughout this paper, we use \textit{target speaker identification} to refer to the combined task---closely related to speaker tracking---of determining when a single registered speaker is active within a continuous audio stream, and \textit{speaker verification} to refer to the pairwise embedding comparison used to make that determination.

Hearing aids process audio in a streaming manner with extremely low latency, as listeners can perceive degradation of quality with latencies as low as 10 ms \cite{lelic2022hearing,stone2003tolerable}. Currently available online diarization systems operate with a 500--1000 ms latency \cite{xue2021online,9414371,coria2021overlap}, and no diarization-based approach can meet the sub-10 ms requirement of the audio playback path. For this reason, a signal-based approach is described here, in which the start point and end point of a target speaker's speech are identified in a streaming manner and emitted as a control signal that operates outside of the audio playback path. The audio itself may continue to be processed and amplified at the native sub-10 ms latency of hearing aid hardware, while the identification signal gates or steers a downstream low-latency amplification algorithm such as the system described by \cite{zheng2022low}, which operates with 4 ms latency. The latency of identification therefore determines only how quickly amplification adapts at speaker turns, not the perceptual delay of the audio presented to the listener.

Evaluating the system on a long-form conversational podcast dataset in a streaming manner, we achieved greater than 90\% accuracy in identifying speech segments of the target speaker. The repository is available from the authors upon reasonable request.

\section{Materials and Methods}

To emulate conversational speech while minimizing overlapping speakers, we selected the This American Life Podcast Transcripts dataset \cite{mao2020speech,tal_kaggle}. The selected dataset consists of over 600 episodes of the This American Life podcast, each of which contains conversations of an average of 18 speakers. The dataset includes granular transcript detail including the speaker name, the speech timestamps, and speech content. To maintain consistency across episodes for the purposes of this project we selected Ira Glass, the podcast's host, as the target speaker. Of the available episodes in the dataset, the latest episodes (those numbered 670--702 and recorded in 2019--2020) were utilized throughout this project. Episodes that did not feature Ira Glass were excluded.

We selected two open source diarization systems to test for a baseline DER metric---Pyannote version 2.1.1 \cite{bredin2023pyannote} and LIUM (available from \cite{du2023lium}). Each of these systems has a robust toolkit available to process pre-recorded audio, and in the case of LIUM it was specifically designed to support broadcast media. We tested each model with the This American Life dataset and compared them using the diarization error rate (DER) as calculated by the Pyannote metrics library \cite{bredin2017pyannote}. We utilized each system according to its default protocol published in the respective documentation and modified each as needed for the This American Life dataset. Based on the comparative DER results and its published benchmark with the This American Life dataset \cite{mao2020speech}, the Pyannote diarization system was selected for the purposes of further configuration and hyperparameter tuning.

We then utilized a streaming diarization system called Diart (version 0.8.0) \cite{coria2021overlap}, which is based on Pyannote and uses ReactiveX through the RxPY library to achieve streaming support with 500 ms resolution. We used a grid search to tune its clustering algorithm and speaker detection parameters with the goal of optimizing the DER.

\begin{figure}[!htbp]
  \centering
  \includegraphics[width=0.55\linewidth]{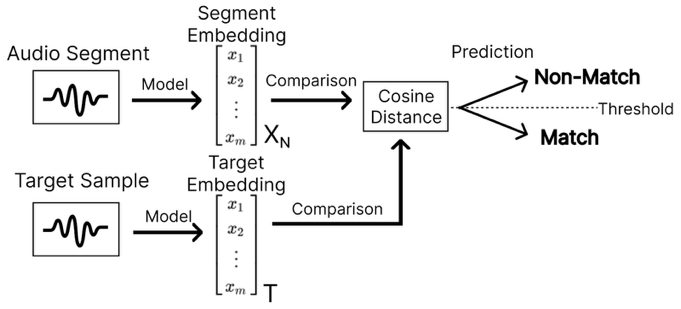}
  \caption{Speaker verification is completed by using a speaker verification model to create a vector embedding of an audio segment and a target speaker's speech segment. The cosine distance is calculated between these embeddings and compared to a threshold value.}
  \label{fig:fig1}
\end{figure}

We selected two well-studied verification models to test: Pyannote (version 2.1.1) \cite{pyannote_embedding} and TitaNet-Large from NVIDIA NeMo \cite{harper2024nemo}. Pyannote offers robust cross-compatibility with the Pyannote-based diarization module in addition to excellent performance, and TitaNet surpasses Pyannote's performance metrics \cite{brydinskyi2024comparison}. In each case, we segmented the source audio from the This American Life dataset using Pyannote diarization. We created target speaker segments from audio segments of varying lengths in which the host Ira Glass was speaking. Finally, we utilized the verification models to create a vector embedding for both the target speaker and a test utterance and calculated their cosine distance (1-[cosine similarity]) as depicted in Figure~\ref{fig:fig1}. We created Receiver Operating Characteristic (ROC) curves to determine the optimum verification model, duration of speech for the target embedding, and the optimum threshold for cosine distance comparison.

We next formulated a streaming diarization and verification system by pairing the Diart diarization system with speaker verification through the ReactiveX framework. While Diart performed well with 500 ms latency, verification required larger audio chunks for performance. We completed further hyperparameter tuning to discern the optimal number of 500 ms chunks for verification.

\begin{figure}[!htbp]
  \centering
  \includegraphics[width=0.5\linewidth]{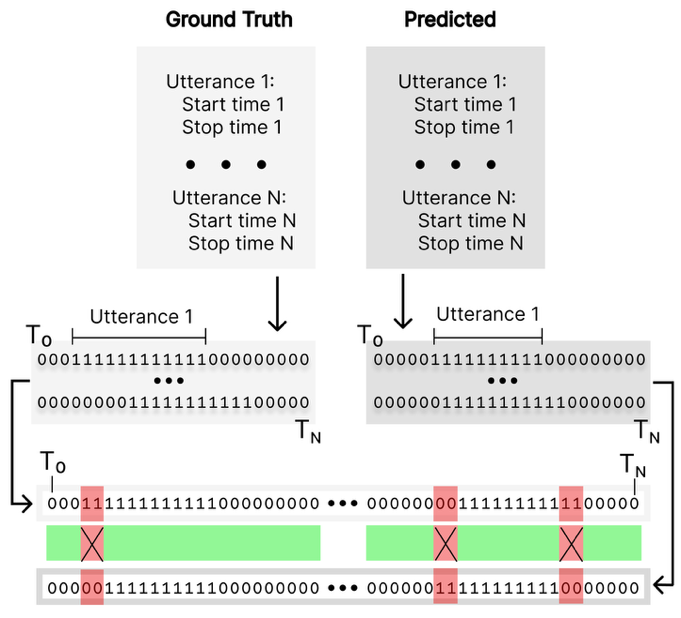}
  \caption{Two boolean vectors are created such that each value represents 0.1 seconds of the podcast. True values in each vector, respectively, represent times when the target speaker spoke according to the ground truth, or when the system predicted target speaker speech. These two vectors were then compared to calculate the system's metrics.}
  \label{fig:fig2}
\end{figure}

Standard calculations for recall, precision, F1 score, accuracy, and specificity were utilized. To perform these calculations, as depicted in Figure~\ref{fig:fig2}, we converted the system output into a binary mask across the length of the This American Life episode to represent predicted times the target speaker was speaking. The continuous timestamp outputs were discretized to a resolution of 100 ms chunks across the episode. The same transformation was also completed using the ground truth files supplied with the dataset before calculating the above scores comparing the predicted and actual masks. For each of these metrics, the mean, median, and standard deviation are reported.

\section{Results}

We developed a real-time pipeline for speaker identification that utilized two distinct stages---speech diarization and speaker verification. These were then optimized based on objective metrics and for maximal interoperability.

We first measured a baseline offline DER metric goal using two open source speech diarization models, Pyannote and LIUM, with their default configurations and a single episode from the \textit{This American Life} dataset. Using this workflow, Pyannote achieved a DER of 0.201, while LIUM's DER measured 0.976. Based on these initial results and its compatibility with the Diart streaming system, Pyannote was selected for further use in the offline evaluation of verification models.

We next evaluated the Diart streaming diarization system and began with a grid search to determine the optimal latency and clustering hyperparameters. Using a single podcast episode and initial parameters based on \cite{coria2023captions}, Diart's configuration space was sampled. The optimal setting identified (delta\_new = 0.895, rho\_update = 0.1, tau\_active = 0.5) yielded a DER of 0.310 for the episode, a 44.9\% reduction relative to the baseline DER of 0.563. Based on these results and its well documented interoperability, Diart was selected for the diarization phase of the pipeline.

\begin{figure}[!htbp]
  \centering
  \begin{subfigure}{0.48\linewidth}
    \centering
    \includegraphics[width=\linewidth]{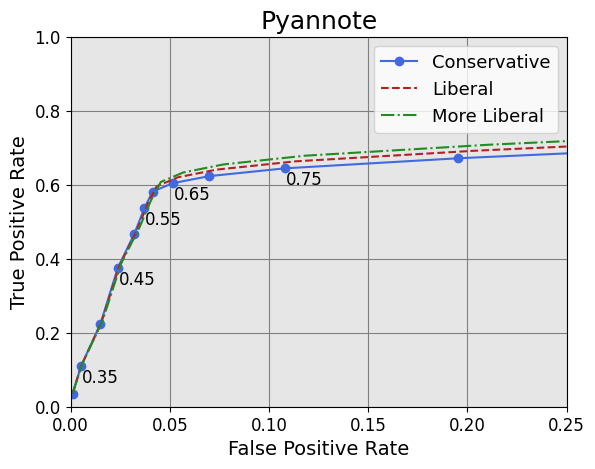}
    \caption{Pyannote}
    \label{fig:fig3_pyannote}
  \end{subfigure}
  \hfill
  \begin{subfigure}{0.48\linewidth}
    \centering
    \includegraphics[width=\linewidth]{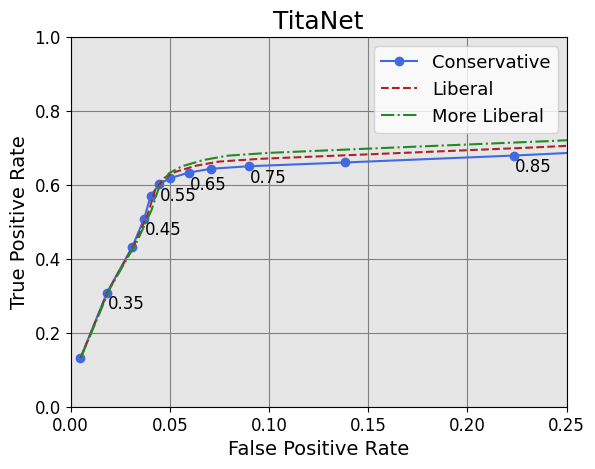}
    \caption{TitaNet}
    \label{fig:fig3_titanet}
  \end{subfigure}
  \caption{Either Pyannote or TitaNet were used to create embeddings from audio segments based on the Pyannote diarization of 23 episodes of the This American Life Podcast. Speaker verification was then performed using various thresholds for the cosine distance. Metrics were then calculated according to Figure~\ref{fig:fig2} and ROC plots were generated. The liberal and more liberal plots were generated by adding 0.5 seconds or 1 second, respectively, to the beginning and end of the predicted target speaker's utterances.}
  \label{fig:fig3}
\end{figure}

Speaker verification models were next evaluated using both Pyannote and NVIDIA's TitaNet-Large. The models created vector embeddings of segments predicted to contain speech. These were then compared using cosine distance to an embedding created from a 57-second segment of speech by the target speaker (Ira Glass). Evaluation was performed across episodes 677--701 for Pyannote and TitaNet, which consisted of 23 episodes as episodes 682 and 683 were not part of the dataset, with cosine distance thresholds ranging from 0.3 to 0.95. The alignment method depicted in Figure~\ref{fig:fig2} was used to calculate true positive and false positive rates. Receiver Operating Characteristic (ROC) plots summarizing these results are depicted in Figure~\ref{fig:fig3}. We define the point of inflection as the knee of the ROC curve: the operating point beyond which further threshold increases produce a disproportionate rise in false positive rate relative to the gain in true positive rate. This point is represented by a threshold of approximately 0.65 in the Pyannote plot and 0.6 in the TitaNet plot. This inflection point is located at the same approximate location in both plots---a false positive rate of 0.05 and a true positive rate of 0.6. Given the similar results between the two models, Pyannote was selected for its superior interoperability with the Diart diarization system.

To account for possible small inconsistencies in timestamps between the ground truth file and those predicted by the model, three alignment variations to Figure~\ref{fig:fig2} were tested. In these alignments, 0.5s was added before and after the predicted speech windows by the target speaker in the ``Liberal'' plot, and 1.0s was added in the ``More Liberal'' plot. The results of these alignment variations are included in Figure~\ref{fig:fig3}.

To assess the effect of target speaker registration length on verification performance, we generated 9 embeddings from target audio segments ranging from 13 to 123 seconds in duration. These were generally created from concatenations of several segments during which Ira Glass was speaking. The Area Under the Curve (AUC) was calculated from ROC curves for each target length compared to the same 23 episodes, using Pyannote for verification and a conservative alignment, which is summarized in Table~\ref{tab:target_auc}. The 57-second embedding was selected for use in downstream testing.

\begin{table}[t]
  \caption{Various target speaker audio segment lengths were created by concatenating known utterances of the target speaker. The embeddings from these segments were each used for speaker verification across 23 podcast episodes and with various threshold values. The AUC was then calculated from ROC plots for each audio segment length.}
  \label{tab:target_auc}
  \centering
  \begin{tabular}{l ccccccccc}
    \toprule
    \textbf{Duration (s)} & 13 & 26 & 38 & 50 & 57 & 66 & 93 & 105 & 123 \\
    \midrule
    \textbf{AUC} & 0.324 & 0.394 & 0.410 & 0.423 & 0.449 & 0.451 & 0.445 & 0.442 & 0.446 \\
    \bottomrule
  \end{tabular}
\end{table}

Lastly, we finalized and evaluated the full system pipeline based on the audio chunk handling described in \cite{coria2023captions}. We utilized Diart for diarization and Pyannote for verification in a streaming, real-time configuration. Based on the operating region identified in Figure~\ref{fig:fig3}, we tested two cosine distance thresholds (0.7 and 0.75) and compared Accuracy, Precision, Recall, F1 Score, and Specificity across 17 episodes that met inclusion criteria. The Diart streaming module streams the audio file programmatically, bypassing the use of speakers and a microphone and yielding consistent results for each file. Thus the system was run a single time on each episode. The results are summarized below in Figure~\ref{fig:fig4} and Table~\ref{tab:metrics_summary}. Diart divides the incoming audio stream into chunks of a specified length, with a minimum of 500 ms. We used a grid search to determine the minimum chunk size and number of chunks that needed to be concatenated to achieve adequate downstream speaker verification and found that 2 audio chunks of length 500 ms provided excellent results with an overall latency of 1 second.

\begin{table}[!htbp]
  \caption{The mean, median, and standard deviation for the metrics plotted in Figure~\ref{fig:fig4}.}
  \label{tab:metrics_summary}
  \centering
  \begin{tabular}{lccc}
    \toprule
    \multicolumn{4}{c}{\textbf{Values for the threshold of 0.70}} \\
    \midrule
    \textbf{Metric} & \textbf{Median} & \textbf{Mean} & \textbf{Standard Deviation} \\
    \midrule
    Accuracy     & 0.93 & 0.91 & 0.05 \\
    F1           & 0.68 & 0.66 & 0.12 \\
    Precision    & 0.91 & 0.86 & 0.14 \\
    Recall       & 0.56 & 0.55 & 0.12 \\
    Specificity  & 0.99 & 0.98 & 0.02 \\
    \midrule
    \multicolumn{4}{c}{\textbf{Values for the threshold of 0.75}} \\
    \midrule
    \textbf{Metric} & \textbf{Median} & \textbf{Mean} & \textbf{Standard Deviation} \\
    \midrule
    Accuracy     & 0.91 & 0.90 & 0.05 \\
    F1           & 0.70 & 0.68 & 0.13 \\
    Precision    & 0.78 & 0.73 & 0.18 \\
    Recall       & 0.68 & 0.65 & 0.10 \\
    Specificity  & 0.96 & 0.95 & 0.05 \\
    \bottomrule
  \end{tabular}
\end{table}

\begin{figure}[!htbp]
  \centering
  \includegraphics[width=0.75\linewidth]{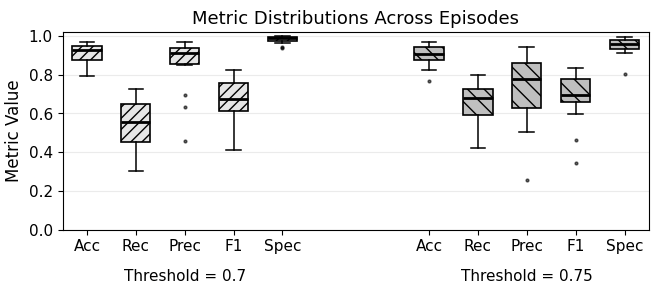}
  \caption{The accuracy (Acc), recall (Rec), precision (Prec), F1 score, and specificity (Spec) were calculated to evaluate the performance of the real-time speaker identification system using two thresholds (0.7 and 0.75) for the cosine distance.}
  \label{fig:fig4}
\end{figure}

\section{Discussion}

The real-time pipeline described here combines speaker diarization and verification to provide a signal-based mechanism to identify a target speaker with very low latency. It was found that accuracy (median 0.91 for threshold 0.75) and specificity (median 0.96 for threshold 0.75) remained high across the episodes tested, indicating overall excellent performance in this dataset. An approximate median of 68\% of target speaker speech time was identified for the same threshold (recall) while limiting false positive prediction time to an approximate median of 22\% of the predicted target speaker speech time (1 $-$ precision). Greater variation was noted in recall and precision than in accuracy or specificity across the episodes. Notably, several consecutive episodes (685--689) yielded a recall that was below the overall average, which likely indicates differences in episode recording conditions. This underscores the importance of maintaining a consistent acoustic setting after registering the target speaker.

Our system enables the registration of an arbitrary speaker prior to the initiation of audio signal processing. We tested a variety of target segment durations (from 13 to 123 seconds) with the Pyannote-based non-streaming system, summarized in Table~\ref{tab:target_auc}. We found improvements in target speaker identification, as measured by AUC, as the target segment length increased up to approximately 1 minute in length. The AUC plateaued with longer segments, likely indicating limited utility in using a very long speaker registration. The use of a 1-minute registration audio segment in practice would provide excellent results while avoiding a significant time burden for the user.

A central challenge in adapting machine learning diarization models for hearing aid applications is the latency imposed by online processing. Diart was able to achieve a DER of 0.310 on a single episode, approaching Pyannote's offline performance (DER = 0.201) while maintaining real-time processing.

Ultimately, our proposed method aims to support an amplification algorithm to augment the target speaker's voice. With Diart's 500 ms minimum and speaker verification's need for multi-chunk aggregation, the total latency achieved by our system was approximately 1 second to identify both the start and end of a target speaker's utterance, which provided excellent performance in the This American Life dataset. Because the identification signal operates outside of the audio playback path, this signal-based approach is not subject to the sub-10 ms latency requirement of hearing aid audio processing---a requirement far below what current state-of-the-art streaming diarization models can achieve.

The This American Life Podcast Transcripts dataset is relatively clean and structured, based on long-form turn-based discussions, and with limited overlapping speech and little background noise. We observed degraded speaker verification performance when background noise was present in the recorded audio. The inclusion of a denoising algorithm, such as the DiffWave-based model described in \cite{zhang2021restoring}, could be beneficial in improving the quality of target speaker detection in more general circumstances.

A latency of 1 second is likely acceptable for sustained conversations, but it may still be suboptimal for rapid exchanges or brief interjections. Additionally, the latency in this algorithm poses an obstacle to the use of diarization for the purpose of parsing overlapping speech in real time. Further work may utilize a hybrid approach combining signal-based methodology with predictive modeling or even novel brainwave-based technology \cite{han2019speaker} to handle overlapping speech. Alternative paradigms may also warrant direct comparison: target-speaker voice activity detection (TS-VAD) \cite{medennikov2020tsvad} detects the activity of an enrolled speaker directly and is well suited to overlapped speech, while target speech extraction methods such as VoiceFilter \cite{wang2019voicefilter} bypass segmentation entirely by separating the target speaker's audio through speaker-conditioned spectrogram masking. These approaches represent promising alternatives or complements to the diarization-plus-verification pipeline described here.

This work has several limitations. The evaluation enrolled a single target speaker, and the baseline diarization comparison and hyperparameter tuning were each performed on a single episode. Results are reported from a single run per episode, without repeated runs or statistical significance testing. Evaluation across a larger pool of enrolled speakers, with tuning and evaluation data kept disjoint, will be necessary to establish broader confidence in the reported performance levels.

In this system, speaker verification with either Pyannote or TitaNet yielded similar results, and in Figure~\ref{fig:fig3} an inflection point can be identified at approximately (0.05, 0.6) in both plots. Using the Pyannote ecosystem for both diarization and verification tasks provided a seamless system, and performed comparably to the use of TitaNet. Future work may explore other diarization or verification models, such as ECAPA-TDNN, which has been shown to perform very well with speaker verification \cite{brydinskyi2024comparison}. Only small differences were noted in the curves in Figure~\ref{fig:fig3} when comparing the various alignment conditions, which suggests that timing errors introduced by diarization have only a small impact on the overall system performance.

Overall, this system offers a promising proof of concept for real-time, low-latency target speaker identification. Such a system could be integrated into modern hearing aids to enable selective amplification without user intervention, with the potential to improve speech intelligibility in group conversations. Our signal-based design decouples the latency of streaming diarization (500 ms or greater in existing systems) from the audio playback path, keeping identification delays out of the audio presented to the listener. This approach preserves the latency budget required for low-latency speech amplification (such as the 4 ms system described in \cite{zheng2022low}) and enables integration into existing hearing aid technology.

\section*{Acknowledgments}
The authors thank the DigitalDX Lab at Arizona State University for their assistance with project design and manuscript revision.

\section*{Generative AI Disclosure}
Generative AI was not used in the formulation or completion of this project.

\bibliographystyle{unsrt}
\bibliography{references}

\end{document}